\documentclass[aps,twocolumn,prb,showpacs]{revtex4}
\usepackage{graphicx}
\usepackage[]{subfigure}
\begin{document}

\title{Pressure dependence of the melting mechanism
at the limit of overheating in Lennard-Jones crystals}

\author{ L.G\'omez, C. Gazza , H. Dacharry, L. Pe\~naranda and A. Dobry}
\affiliation{Facultad de Ciencias Exactas Ingenieria y Agrimensura,
Universidad Nacional de Rosario\\
and Instituto de F\'{\i}sica Rosario, Avenida Pellegrini 250, 2000 Rosario,
Argentina}
\date{\today}
\begin{abstract}
We study the pressure dependence of the melting mechanism of a
surface free Lennard-Jones 
crystal by constant pressure Monte Carlo simulation.
The difference between the overheating temperature($T_{OH}$) and the thermodynamical 
melting point($T_M$) increase for increasing pressure.  
When particles move into the repulsive part of the potential 
the properties at $T_{OH}$ change. There is a crossover
pressure where the volume jump becomes pressure-independent.
 The overheating limit  
is pre-announced by thermal excitation of  big clusters of
defects. The temperature zone where the system is dominated by
these big clusters of defects increases with increasing pressure.
Beyond the crossover pressure we find that excitation of defects and 
clusters of them start at the same temperature scale related with $T_{OH}$.
\end{abstract}
\maketitle

\section{Introduction}
Melting is a dramatic phase transition where the translation
symmetry present in the solid phase is destroyed at the transition point.
As a paradigm of a phase transition in condensed matter physics, its
microscopic mechanism has been studied from the beginning of the solid
state theory~\cite{Dash}.

Early theories relate melting with an intrinsic elastic
instability of the crystal. According to a stability criteria
established by Born~\cite{Born}, melting takes place when an
elastic shear modulus vanishes and the crystal looses its ability
to resist shear. This was interpreted as due to the softening of a
transverse phonon mode with, as a consequence, the homogeneous
breakdown of the crystalline order. This notion of crystalline
instability refers only to the properties of the crystalline phase
but a complete theory of melting should also refer to the
properties of the liquid phase.

Therefore, modern theories invoke thermal excitation of extended
defects such as dislocation lines~\cite{Bura,Lund}. Pairs of
dislocations can be thermally excited at temperatures within the
range of typical solid phases. The presence of dislocations lowers
the cost in energy to create an additional dislocation; if the
energy reduction is strong enough, it could lead to an avalanche
of dislocations resulting in a first order phase transition.

So far, two different ways of melting has been recognized in the
literature \cite{Cahn}. One related to the thermodynamical melting starting at
the surface or extrinsic inhomogeneities of the samples, and the other one with 
regards to the limit of the mechanical stability of the crystal 
which takes place when surface is avoided and the solid could be overheated 
above the equilibrium melting point. Overheating has been experimentally achieved by
coating a solid with another material of higher melting
point~\cite{Daeges}.

 As no extrinsic inhomogeneities are present,thermal excitations of defects 
play an essential role to initiate the melting process. In this work, we will 
focus on this type of melting.

Although experimental analysis of the microscopic dynamics near
melting
 is very difficult, numerical simulations have recently
shown that correlated clusters of defects thermally excited play a
central role in this process at the limit of overheating~\cite{JinPRL}. 
Moreover in a previous work we have related clusters of defects to  dislocation 
lines assumed in phenomenological theories~\cite{Gomez}. However, the situation is
far from being solved and continues to  puzzle  condensed-matter theorists.

An alternative strategy to look for the melting origin could be to
change an external parameter and to follow how this parameter
affects the melting. The obvious choice of this parameter is the
external pressure acting on the solid. Moreover, the pressure
dependence of the melting temperature has  a practical interest
{\it per se} and  the determination of high pressure melting
curves is a subject of many investigations~\cite{HighPmelt,Belonoshko}.

The purpose of this paper is to discuss the effect of pressure on
the melting properties of a surface free crystal by means of a computationally
intensive  Monte Carlo (MC) calculation using a constant pressure
algorithm.

The paper is organized as follows. In Sec. II we analyze some
thermodynamic properties at overheating limit. In Sec. III we look at the
defect structure preceding the overheating limit. Section IV contains our
conclusions and discussion.

\section{Phase diagram of the Lennard-Jones system}

Our simulations have been performed on a cubic box of 2048
particles~\cite{size} interacting via a Lennard-Jones (LJ)
potential written as $V(r)=4\epsilon [(\frac{\sigma
}{r})^{12}-(\frac{\sigma}{r})^{6}]$. To guarantee that the
potential goes smoothly to zero at distance greater 
than the cutoff ($r_c$), a cutoff region from $0.95 r_c$
to $r_c$ was used following previous work \cite{Agrawal, MorrisSong}.
Therefore, we can compare our results for the properties at 
the overheating limit with ones previously obtained at the equilibrium melting point.
Moreover we fix the cutoff $r_c=2.1 \sigma$ which has been shown to be greater enough to assure convergence of the results\cite{MorrisSong}.

The energies are measured in units of
$\epsilon$, the distances in units of $^3\!\!\sqrt{4}\sigma$ and
the pressure in unit of ${\epsilon}/{4 \sigma^3}$. The temperature
is reported in units of $\epsilon/k_B$. As a concrete example we
can take the parameters for argon
($\sigma=3.4~\AA,~\epsilon=0.0104 eV$) where our units are: 
$5.397\AA$ for the distances, $120.64 K$ for the temperatures and
$10.5\times10^6~Pa$ for the pressures. Our units of pressure
differ by a factor $\frac{1}{4}$ from those used in previous works\cite{Agrawal, MorrisSong}.
Moreover, our unit of volume differ by a factor $4$.

We emphasized that the use of periodic boundary conditions (PBC) suppresses
 surface effects and  the phase transition temperature we find 
does not correspond to the thermodynamical melting point but to the limit of 
overheating of the crystal~\cite{JinPRL}.

\begin{figure}
\centerline{\includegraphics[width=8cm]{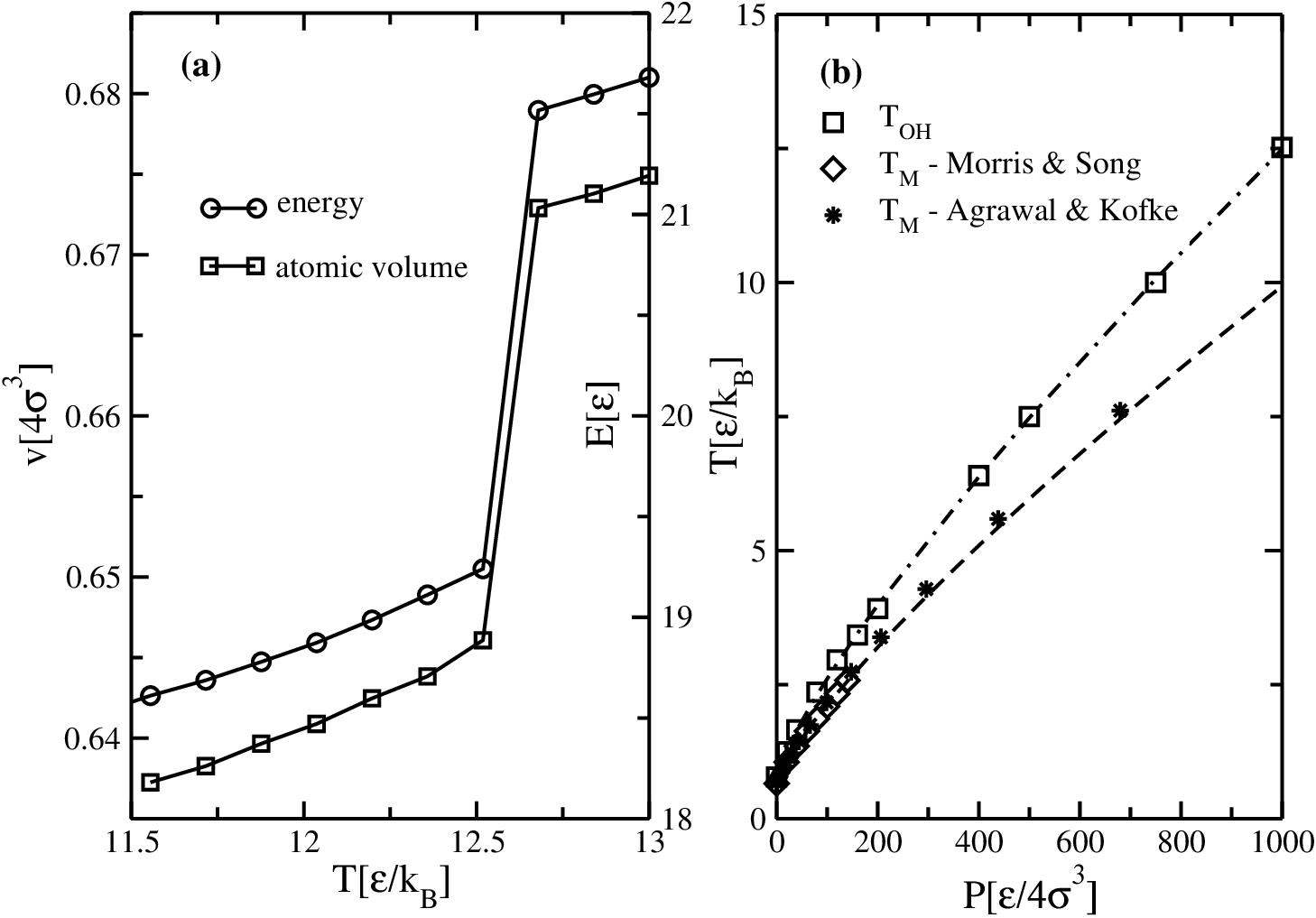}}
\caption{\label{p1000} (a) atomic volume (square, left scale) and
internal energy  (circles, right scale) vs. the temperature ($T$)
for pressure $P=1000$ (b) the overheating temperature ($T_{OH}$) as a function
of the pressure . We also include the melting point $T_M$ as obtained in 
Ref. \onlinecite{MorrisSong} (open diamond) and 
from Ref. \onlinecite{Agrawal} (starts).
Dashed (dot-dashed) lines correspond to the power law 
fitting given in eq. (1) (eq. (2)) shown in the text.}
\end{figure}

Let us discuss how we constructed the phase diagram of the system.
For a fixed pressure at a given temperature, we have averaged the
internal energy and the volume over the MC runs. Both the particle
coordinates and the volume of the cell are updated in each MC
step~\cite{Allen}. Heating is done in a step-by-step procedure
starting at low temperature with the particles at the sites of a
fcc perfect lattice. The calculation of the jump for different
properties (volume, energy, etc.) at the limit of overheating
 is a difficult task in
MC simulation, because of the long CPU time necessary to achieve
equilibrium values in this intermediate region. We have densely
partitioned the temperature interval and run $\sim10^5$ MC steps
in this zone. In Fig.~\ref{p1000}(a) we show a typical results of
the evolution of the energy and the atomic volume ($v$) for $P=1000$ with 
the temperature. The abrupt jump is related to the limit of overheating. 
We have observed that both the snapshot of the particle
positions as well as the radial distribution function (RDF) above
this jump do not show any indication of the crystalline order. By
locating the temperature where this jump takes place at different
pressures we construct the figure ~\ref{p1000}(b) where we show the
overheating limit($T_{OH}$) as a function of pressure. For comparison, 
we have included the results for the equilibrium melting point($T_{M}$) 
extracted from Table II of Ref. \onlinecite{MorrisSong}, obtained by using 
molecular dynamics (MD) coexistence simulations. 
We take from this work the results obtained using the same smooth cutoff 
$r_c=2.1 \sigma$ as in our MC simulation. 

The two set of values could be fit following the corresponding next laws:
\begin{eqnarray}
T_M=0.036 \,P^{0.805} + 0.66
\label{fittm}
\end{eqnarray} 
\begin{eqnarray}  
T_{OH}=0.044 \,P^{0.806} + 0.78
\label{fittoh}
\end{eqnarray}
Note in figure ~\ref{p1000}(b) that even at $P=0$, the overheating phenomena is observed. While the difference between $T_M$ and $T_{OH}$ increase with pressure, the exponent of the pressure is quite similar, $\sim \frac{4}{5}$. 
In the same figure we have added $T_{M}$ datas (start symbols) from Ref. \onlinecite{Agrawal}, where the Gibbs-Duhem integration molecular simulation thechnique was used. We can see the 
agreement with the extrapolation of eq. (\ref{fittm}) even in the high pressure regime\cite{Belonoshko1}.

Let us analyze this exponent at the light of previous results valid in the regime of high pressure\cite{Hansen}. When particle are close enough, the repulsive  
part of the potential $\frac{1}{r^12}$ dominates and a power law has been deduced ($P\beta^{\frac54}=Const$). 
All this, indicate that the low pressure zone could be neglected in the scale we are 
presenting the phase diagram.
 
A microscopic understanding of the differents multiplicative constant shown in the 
eq. (\ref{fittm}) and eq. (\ref{fittoh}) could be explained with the following argument. 
 While no extrinsic nucleation center as surface or defects are present in 
our simulation system, in order to break the crystalline order thermal activation of 
clusters of defects are necessary (see next Section). However, as pressure increase, more 
thermal energy is needed because particles are strongly compacted. This is different 
than the process  involved in the thermodynamical melting where a liquid phase starts at 
the surface and pressure is less effective to increase the transition temperature.    

The Born criteria has been recently reobtained for homogeneous
lattices under an arbitrary but uniform external
load~\cite{Wang93-95,WangPhysA}. In the case of hydrostatic
pressure the generalized stability criteria are given by:
\begin{eqnarray}
\label{BT}
B_T=(C_{11}+2 C_{12}+P)/3 > 0\\
\label{G}
G=C_{44}-P > 0\\
\label{Gp}
G^{\prime}=(C_{11}-C_{12}-2 P)/2 > 0
\end{eqnarray}
where $C_{11}$, $C_{12}$ and $C_{44}$ are the three different
elastic constants. The $B_T$, $G$ and $G^{\prime}$ are the
generalized bulk modulus and the two different shear modulus. We
have obtained the elastic constants by means of the harmonic
approximation ~\cite{Bruesch} on the Lennard-Jones fcc crystal 
with interactions up to the third nearest neighbour and 
variable $v$. This procedure gives the elastic constants at $T=0$
under pure dilation or compression. Note that for a general value
of $v$, the crystal is not in equilibrium under the action of the
interacting forces and the Cauchy relation $C_{12}=C_{44}$ is not
in general satisfied.
\begin{figure}
\centerline{\includegraphics[width=8cm]{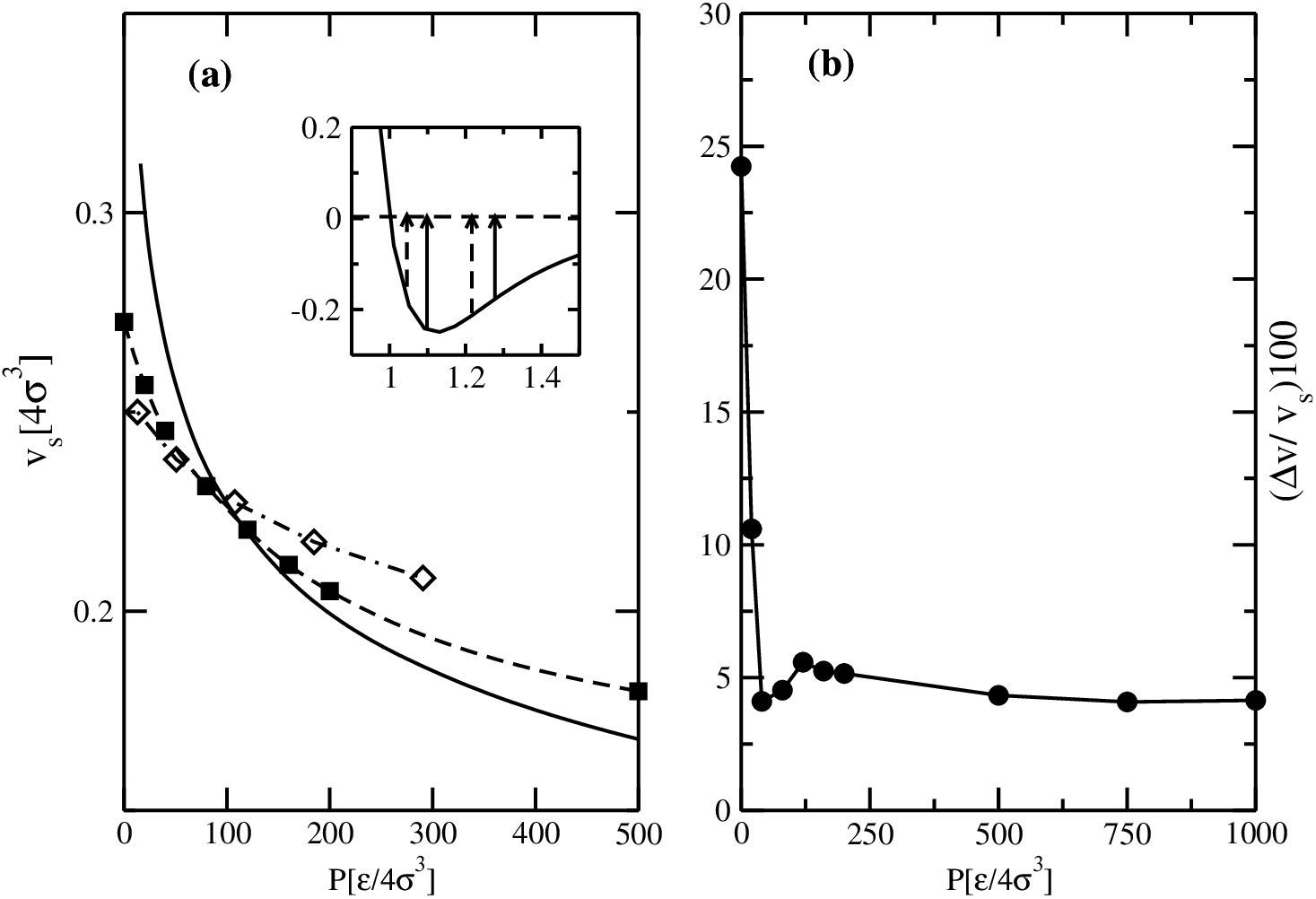}}
\caption{\label{phdiag} (a) The atomic volume of the solid ($v_{OH}$)
at the overheating limit as a function of $P$ (dashed line, square symbols).
The solid line shows the volume where $G^\prime$ vanishes at $T=0$. 
We also include the volume at the melting 
point ($v_M$) extracted from Ref. \onlinecite{Agrawal}
(dot dashed line, diamonds). The inset shows the volume dependence of the
cohesive energy of an ideal fcc crystal. The dashed (solid) arrow signal the
thermal energy to melt (overheat) the crystal.
(b) The percentage jump of the atomic volume at the overheating limit.}
\end{figure}
For each pressure we look for the value of $v$ where some of the
module of Eq.(\ref{BT}-\ref{Gp}) vanish. At $P=0$, $B_T$ is the
first to be vanished, i.e. for smaller $v$, than the two shear
modules. This instability take place at $v=0.315$ higher than the
melting detected by MC simulation ($v\sim0.273$). From $P>15$ the
shear modulus $G^{\prime}$ is  the first to become zero. This $v$
signals the volume at $T=0$ where the system could not resist
shears. In Fig.\ref{phdiag}(a) we show these values (solid line)
together with the value of $v$ corresponding to the overheating ($v_{OH}$)
obtained by MC simulation (square symbols and dashed line).

We have also included in the figure the volume at the melting point($v_M$). The density($\rho$)-temperature($T_M$) relationship were extracted from Ref. \onlinecite{Agrawal}. The specific volume is $v=\frac{1}{\rho}$ and the pressures at
melt were obtained from $T_M$ using eq. (\ref{fittm}).  

The three curves intersect at a pressure which separates
 two regimes. At this pressure, $P_c\!\sim100$, $v_{OH}$ is
approximately the value where a perfect fcc lattice is in
equilibrium under the action of the interacting forces. Therefore
the system is dilated at melting for $P<P_c$. For $P>P_c$, it is
compressed. Note that the stability criteria given by Eq.(\ref{Gp}) gives an
overall correct estimation for the overheating volume. The fact that the 
vanishing of $G^\prime$ shear module gives a good estimation for the superheating volume when
solid expands has been recognized in previous
works~\cite{JinPRL,WangPhysA,Sorkincondmat}. Note however that in a recent
study of amorphization under decompression it has been shown
that the condition $G^\prime=0$ and the softening of a shear
phonon signal the critical pressure which destroys the crystalline
order~\cite{Jagla}, only at $T=0$.

Note in addition, that  $v_M$ is smaller (greater) than $v_{OH}$ 
for the dilated (compressed) system. This is consistent with the 
overheating phenomena found for all pressure.
Looking at the inset it is easy to note two different behaviors around the equilibrium 
volume ($v_e$). For volumes bellow the $v_e$ the thermal energies necessary to destroy 
a crystal are lower in the overheating case. And in contrary, for volumes above 
$v_e$ the thermal energies are lower in the melting case (see the inset of fig. 
\ref{phdiag}).
 
The crossover from dilation to compression is also seen in the
percentage jump of the atomic volume at the melting $\Delta
v/v_{OH}$($\Delta v$ is the difference between the liquid and solid
volumes per particle). In Fig.~\ref{phdiag}(b) we show this value
obtained from MC simulation. We can see that for
$P\!>>\!P_c$,  $\Delta v/v_{OH}$ is reduced and becomes almost
pressure independent. This behavior could be understood from the relation 
valid in the high pressure regime. 
From Eq. (18) of Ref. \onlinecite{MorrisSong} we see that the specific volume of the 
solid ($v_S$) and the liquid ($v_L$) scale with the same exponent of the 
temperature:
\begin{eqnarray}
4 v_L=C_1 \beta^\frac14\nonumber\\
4 v_S=C_2 \beta^\frac14  
\end{eqnarray}
with $\beta=\frac{1}{k T}$, $C_1=1.23$ and $C_2=1.18$. The relative jump of 
the volume is:
\begin{eqnarray}
\frac{\Delta v}{v_{s}}=\frac{C_1}{C_2}-1=0.042 
\end{eqnarray}
which is $T$ and $P$ independent and of the order of magnitude of 
the value found in the MC simulation at high pressure as can be seen in 
Fig. \ref{phdiag} (b)
 
Microscopically, the structure of the solid near the overheating limit is not 
the same above and below $P_c$. Indeed, the jump percentage  of $v$ becomes very 
small above $P_c$ signalling that the structure of the solid below $T_{OH}$ has 
a lot of defects and could be easily destroyed. We will return to this
point in the next section where the evolution of the defect structure will be analyzed. 

\begin{figure}
\centerline{\includegraphics[width=8cm]{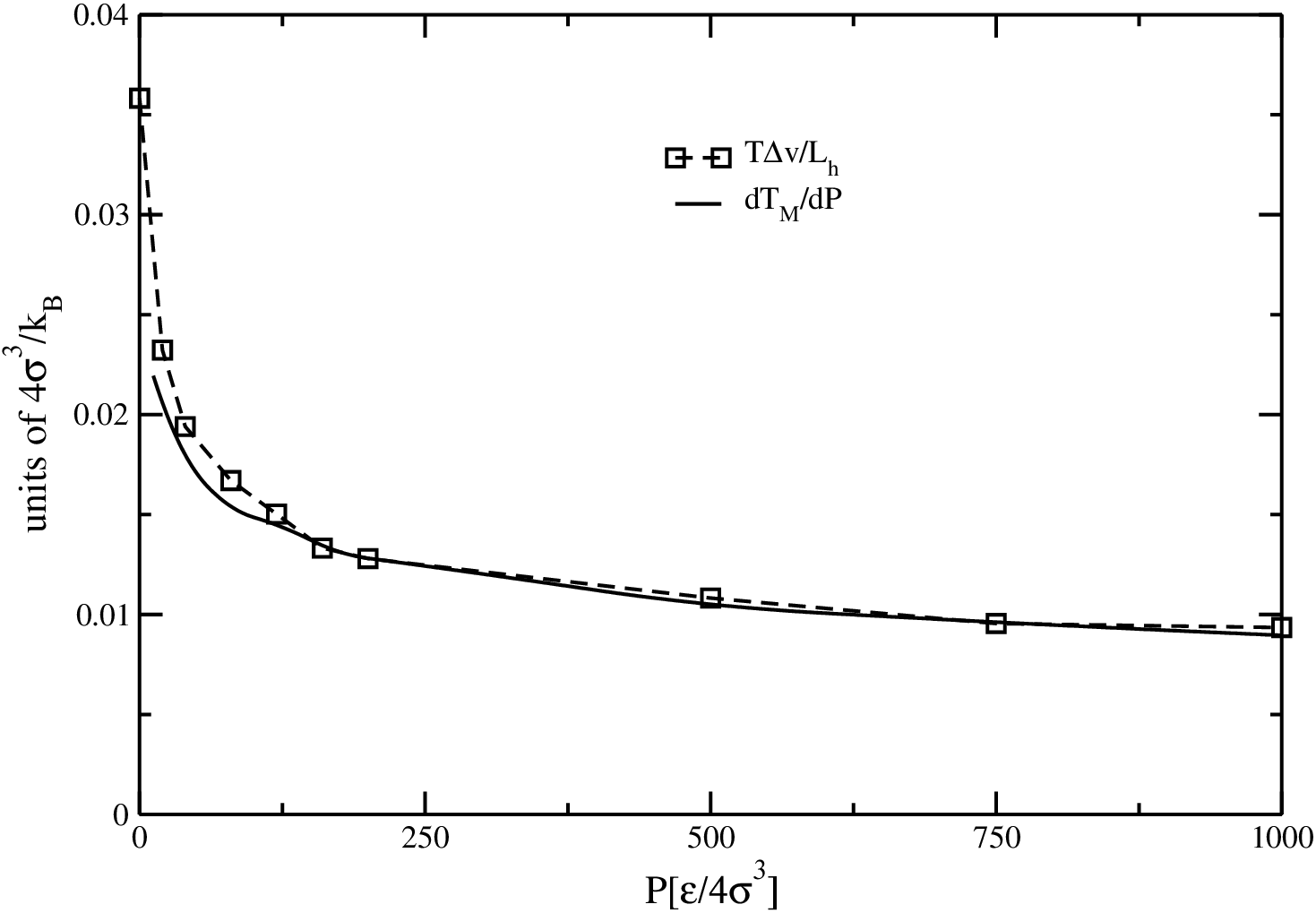}}
\caption{\label{clauscla} The two terms of the Clausius-Clapeyron
equation evaluated numerically. The solid line corresponds to
$\frac{dT_{OH}}{dP}$ evaluated from Eq. \ref{fittoh}. Squares correspond to
$\frac{T_{OH} \Delta v}{L_h}$ where the latent heat $L_h$ is obtained
from the jumps of the internal energy and the volume ($\Delta v$).}
\end{figure}

We now look at the two term of the Claussius-Clapeyron equation obtained from our 
numerical results. Note that the validity of this equation is not guaranteed a priori 
because we are calculating the properties at the overheating limit where the 
free energy of the liquid and solid phase are not equal. It is an interesting point 
to analyze how different are the properties at $T_{OH}$ from the ones predicted 
assuming coexistence between the two phases. 
We have calculated the latent heat of the
transition defined as $L_h=\Delta u+ P \Delta v$, where $\Delta u$
is the jump of the internal energy. In Fig.~\ref{clauscla} we
compare the two terms of the Clausius-Clapeyron equation $T\Delta
v/L_h$ with $\frac{dT_{OH}}{dP}$ obtained by derivation of
the fitting law Eq. \ref{fittoh}. Striking the two term coincide at high pressure. 

\section{Structure of defects preceding the overheating limit}
We have recently found~\cite{Gomez} a premelting temperature at
$P=0$ where the number of thermally activated defects increases
dramatically and defects start to group in clusters. Near $T_{OH}$
 only one cluster across the whole system survives,
breaking the crystalline order. In this work we analyze how this
feature evolves with the external pressure.

We define a defect as a particle with coordination number (CN)
different from $12$, which is the number of nearest neighbors (NN)
in an ideal fcc lattice. This CN, called $C_{NN}$, is obtained by
counting the number of particles around a given particle up to a
cutoff radius.  This cutoff is chosen for each pressure as the
value where the radial distribution function has its first minimum
at low temperature, i.e. the value of the distance between the
maxima of the first and the second neighbors.

\begin{figure}
\centerline{\includegraphics[width=8cm]{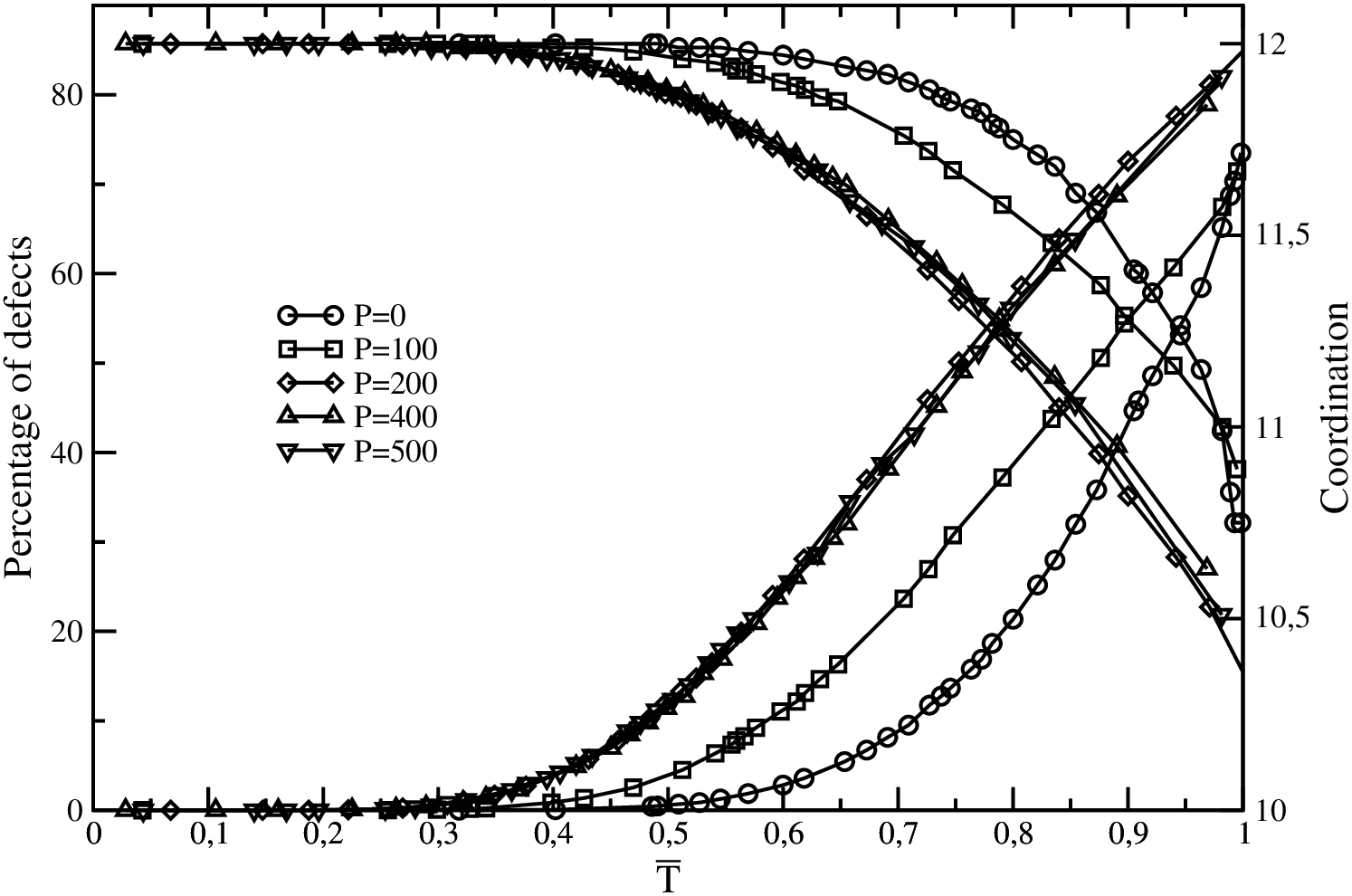}}
\caption{\label{CNvsT} The percentage of defects (left scale) and
the mean coordination number (right scale) as functions of the
reduced temperature for different pressures.}
\end{figure}

To compare our results for different pressures we normalize the
temperature scale to the value of $T_{OH}$ corresponding to each
pressure ($\overline{T}=T/T_{OH}$). In Fig.~\ref{CNvsT} we show the
coordination number mean value and the percentage of defects as a
function of $\overline{T}$ for different pressures. We see that
the reduced temperature where a great number of defects are
activated and the crystalline order is perturbed, decreases by
increasing pressure. For example for $P=0$ we should go up to
$70\%$ of $T_{OH}$ to have $10 \%$ of defects whereas at $P>P_c$ we
achieved the same quantity of defects by going only up to $50\%$
of $T_{OH}$. The decrease of the coordination number shows a similar
behavior. These facts are consistent with the small jump of the
volume  between the two phases found in the previous section.

Moreover, it is seen in Fig.~\ref{CNvsT} that going beyond the
crossover pressure $P_c$ the curves almost coincide. 
When the repulsive part of the potential dominates the system could be 
connected with a hard sphere model with a temperature dependent radius
\cite{hansenverlet}.
As we normalize the temperature scale at the $T_{OH}$, the coincidence of 
the defect structure at  high pressure is a manifestation of the equivalence 
of the models to an effective hard sphere model.  

Like we stated in the introduction, modern theories associate the
melting process with a proliferation of dislocation lines. This
means that defects would have to be correlated near the melting
point. To analyze this correlation, we have grouped the defects
into clusters by the following methodology. We start from a given
defect and search for new defects up to the cutoff distance. For
each of these new defects, we undertake the same procedure. We
iterate this process up to completion of a cluster of connected
defects. Then we take a new defect disconnected to all the
previous ones and develop the same procedure. At the end of this
process we separate our set of defects in $N_{cl}$ clusters which
are disconnected between them. The defects within a cluster are
neighbors to each other but they are not connected to the defects
of another cluster. 

\begin{figure}
\centerline{\includegraphics[width=8cm]{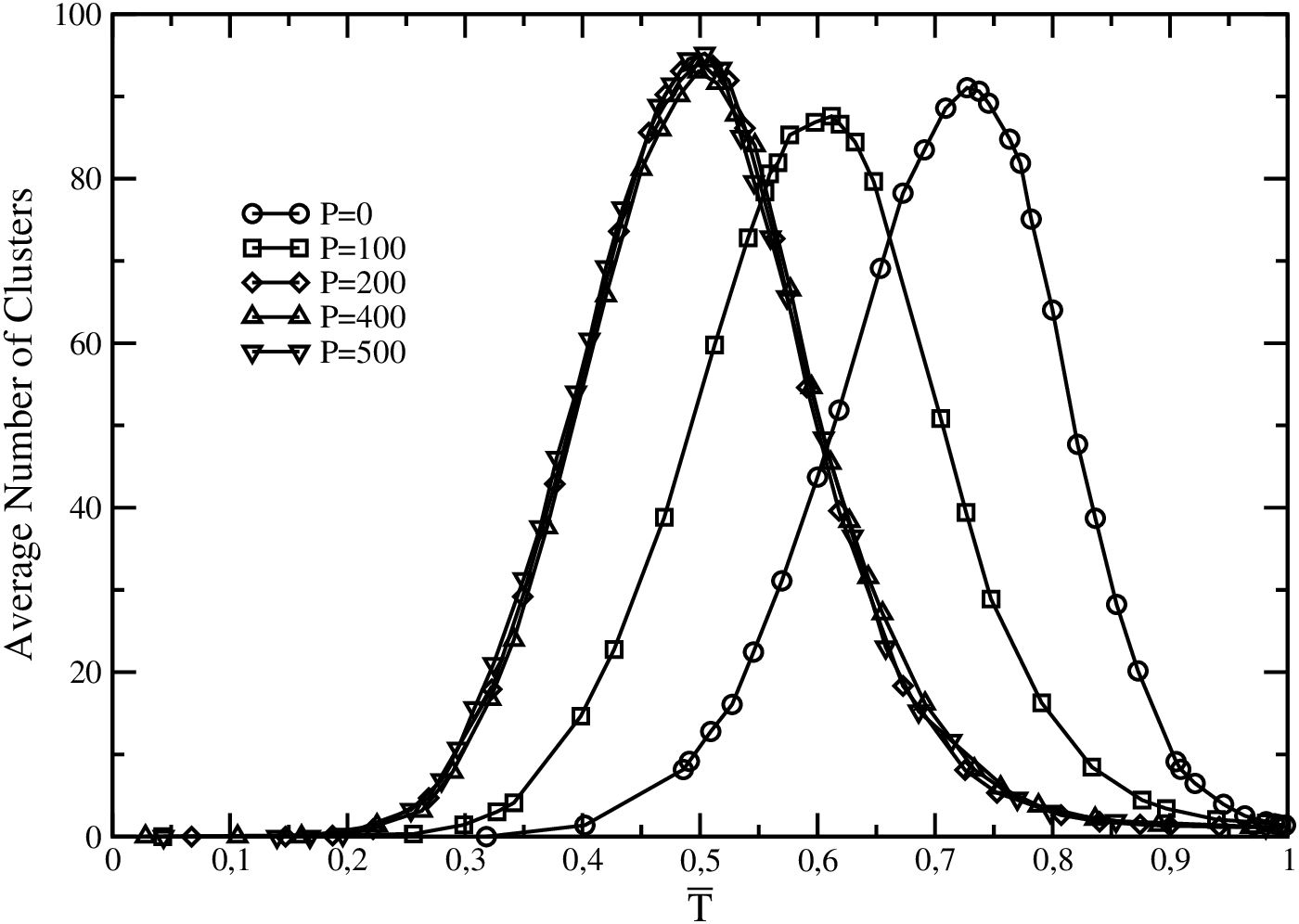}}
\caption{\label{NclvsT} The average number of clusters as a
function of $\overline{T}$  for different pressures.}
\end{figure}

In Fig.~\ref{NclvsT} we show the mean value of $N_{cl}$ as a
function of $\overline{T}$. The decrease of this quantity above a
given $\overline{T}$ (called $\overline{T}_{pm}$) should be
compared with the increase of the number of defective particles
shown in  Fig.~\ref{CNvsT}. These facts indicate that the clusters
are becoming bigger and that the defects correlate among them for
$\overline{T} > \overline{T}_{pm}$. $\overline{T}_{pm}$ decreases
with increasing pressure and becomes constant at high enough
pressures. The premelting zone increases as we already emphasized
previously. This could be important to detect experimentally
excitations of dislocations lines in the study of melting of
materials under pressure.

For low pressures we see that the curves  move
away from the $P=0$ curves and for high pressures they notably
 coincide with each other. All of these seem to point out that
the same mechanism makes  the crystal melt at these values of
pressure. Note that in a phenomenological theory  the crystal
melts at a fixed density of dislocation lines at different
pressures~\cite{BuraJAP}. At this stage, we cannot give a
definitive statement on this prediction but it is suggestive that
the tails of all the curves of Fig.~\ref{CNvsT} correspond to only
one cluster. This means that all the defects of the system are
interconnected to each other.

Finally, let us notice that Fig.~\ref{CNvsT} and Fig.~\ref{NclvsT}
show not only the same mechanism of melting at different pressures
but also the same structure of thermal excited defects at these
pressures.

\section{Conclusions and discusion}
We have studied the effect of pressure on the overheating properties
of a crystalline solid. 
We have compared the properties at the overheating limit with the ones
at the thermodynamical melting where the solid and liquid could coexist.
We have connected the thermodynamic
properties with the microscopic dynamics near the overheating limit.
 We have
taken as a representative example an fcc crystal whose particles
interact by a Lennard-Jones potential. We have shown that it is
possible to obtain equilibrium properties of both phases in the
overheating zone by constant pressure MC simulation.

Important conclusions of our study are:\\
(a) The difference between $T_{OH}$ and $T_M$ increase with increasing pressure.\\
(b) A similar power law fit both the pressure dependence of $T_{OH}$ and $T_M$. \\
(c) Volume expansion due to thermal effects and volume compression
due to the external pressure balance at a given pressure. This
pressure signal a crossover between two regimes. The volume jump
between the two phases becomes rather small($\sim 5\%$)well above
the crossover pressure.\\ 
(d) The previous result is interpreted at microscopic level
as due to the existence of
some liquid-like structures in the solid phase . These structures
appear as defects in the crystalline phase. Indeed our results
show that a great number of defects are activated above a
premelting temperature. On the reduced temperature scale, this
temperature decreases by increasing pressure and  saturates at
high pressures.\\
(d) The premelting temperature also signals the beginning of a
temperature range where defects correlate between them grouping in
clusters. The number of clusters decreases as the overheating limit
approaches from low $T$ by cluster collapsing, leaving only one
cluster near the overheating limit.\\

Our results indicate that experimental detection of premelting
behavior, which has been elusive up to now, could be easier
detected in high-pressure studies of the melting.

We acknowledge useful discussions with L. Burakovsky and H. T. Diep.
We thank A. Belonoshko for sending the datas of his Ref. \onlinecite{Belonoshko1}. 
C.G. thanks to Fundaci\'on Antorchas for partial support.

\end{document}